# Complex Patterns in Reaction-Diffusion Systems: A Tale of Two Front Instabilities

Aric Hagberg
Program in Applied Mathematics
University of Arizona, Tucson, AZ 85721

Ehud Meron
Arizona Center for Mathematical Sciences
Department of Mathematics
University of Arizona, Tucson, AZ 85721

May 16, 1994

## Abstract

Two front instabilities in a reaction-diffusion system are shown to lead to the formation of complex patterns. The first is an instability to transverse modulations that drives the formation of labyrinthine patterns. The second is a Nonequilibrium Ising-Bloch (NIB) bifurcation that renders a stationary planar front unstable and gives rise to a pair of counterpropagating fronts. Near the NIB bifurcation the relation of the front velocity to curvature is highly nonlinear and transitions between counterpropagating fronts become feasible. Nonuniformly curved fronts may undergo local front transitions that nucleate spiral-vortex pairs. These nucleation events provide the ingredient needed to initiate spot splitting and spiral turbulence. Similar spatio-temporal processes have been observed recently in the ferrocyanide-iodate-sulfite reaction.

# 1 Introduction

Labyrinthine patterns, spot replication, and spiral wave turbulence are all examples of complex spatial and spatiotemporal patterns exhibited in reaction-diffusion systems. Labyrinthine patterns have been observed in a variety of gradient systems including garnet layers [1], ferrofluids [2], and block copolymers [3, 4]. Recently they also were found in the bistable FIS (ferrocyanide-iodate-sulfite) reaction [5]. This nongradient system also exhibits spot replication and spiral turbulence [6].

Much of this phenomenology can be understood in terms of two key front instabilities: an instability to transverse perturbations reminiscent of the Mullins-Sekerka instability in solidification fronts [7], and a nonequilibrium analog of the Ising-Bloch transition in ferromagnets that we call a Nonequilibrium Ising-Bloch (NIB) front bifurcation [8, 9]. At a NIB bifurcation, a stationary "Ising" front bifurcates to a pair of "Bloch" fronts propagating in opposite directions. The coexistence of counterpropagating fronts is a nongradient effect that lies at the heart of the complex spatio-temporal processes. Beyond the transverse instability and deep in the Ising regime labyrinthine patterns may develop when small disturbances on planar Ising fronts grow and fill the system through fingering and tip splitting. In the vicinity of the NIB bifurcation intrinsic perturbations, such as curvature, may drive dynamic transitions between the two counterpropagating fronts and lead to spot splitting and spiral breakup [10].

To study these instabilities we consider a nongradient doubly diffusive reaction-diffusion model for two scalar fields $u$ and $v$:

$$u_t = u - u^3 - v + \nabla^2 u, \qquad (1.1a)$$
$$v_t = \epsilon(u - a_1 v - a_0) + \delta \nabla^2 v. \qquad (1.1b)$$

This type of model has been studied in the context of nerve conduction (with $\delta = 0$) and chemical reactions [11, 12], semiconductor etalons [13], and many other systems [14]. For $\epsilon \gg 1$ it reduces to a gradient system like those studied in the context of phase separating systems [15, 16].

Equations (1.1) contain four parameters: $\epsilon$, the ratio of the time scales associated with the two fields, $\delta$, the ratio of the diffusion constants, and two parameters, $a_1 > 0$ and $a_0$, that determine the number and type of homogeneous steady states. These states are found by the intersections of the nullclines $v = u - u^3$ and $v = (u - a_0)/a_1$. For this study we always choose $a_1$ and $a_0$ so there are three intersections, each on a different branch of the cubic nullcline. The intersections on the outer branches represent stable solutions which we will call $(u_+, v_+)$ for the positive branch "up" state, and $(u_-, v_-)$ for the negative branch "down" state. When $a_0=0$ equations (1.1) have an odd symmetry and $(u_+, v_+) = -(u_-, v_-)$.

We focus in this paper on the regime $\epsilon/\delta \ll 1$ where a singular perturbation analysis of (1.1) can be performed. In section 2 we review the derivation of the Ising-Bloch front bifurcation line in the $\epsilon-\delta$ parameter plane [17, 18]. In section 3 we investigate the stability of planar Ising and Bloch fronts to transverse perturbations. Implications of the two instabilities on the formation of labyrinthine patterns, spot splitting, and spiral turbulence are studied in section 4. Section 5 contains a description of the numerical methods used in the simulations and Section 6 concludes by connecting this work to a few others.

## 2 The NIB Bifurcation

In addition to the two stable homogeneous solutions, equations (1.1) also admit front solutions connecting regions of $(u_+, v_+)$ and $(u_-, v_-)$. The stability and type of these fronts depends upon the size of both $\epsilon$ and $\delta$. Decreasing from large $\epsilon$, a single Ising front solution bifurcates to form two counterpropagating Bloch fronts.

To study this bifurcation we consider one-dimensional front solutions propagating at constant speeds and connecting the up state at $-\infty$ to the down state at $+\infty$. These solutions satisfy

$$\mu u_{\zeta\zeta} + c\delta\mu u_\zeta + u - u^3 - v = 0, \tag{2.1a}$$

$$v_{\zeta\zeta} + cv_\zeta + u - a_1 v - a_0 = 0, \tag{2.1b}$$

where we rescaled space and time according to

$$z = \sqrt{\mu}x, \ \tau = \epsilon t, \ \mu = \epsilon/\delta \ll 1, \tag{2.2}$$

and introduced the traveling frame coordinate $\zeta = z - c\tau$. Front solutions of (2.1) can be separated into two parts pertaining to distinct regions: outer regions, away from the front, where both $u$ and $v$ vary on a scale of $\mathcal{O}(1)$, and an inner region, including the front, where $u$ varies much faster than $v$. In the outer regions the derivative terms in (2.1a) can be neglected leading to the solutions $u = u_\pm(v)$ of the remaining cubic relation $u - u^3 - v = 0$. Using these forms in (2.1b), and setting the front position, $u = 0$, at the origin, $\zeta = 0$, we obtain closed equations for $v$,

$$v_{\zeta\zeta} + cv_\zeta + u_\pm(v) - a_1 v - a_0 = 0, \tag{2.3}$$

with $u = u_+(v)$ when $\zeta < 0$ and $u = u_-(v)$ when $\zeta > 0$. To simplify, we choose $a_1$ large enough so that $|v| \ll 1$ and the branches $u_\pm(v)$ can be approximated by the linear forms $u_\pm(v) = \pm 1 - v/2$. We then obtain the following linear boundary value problems for the two outer regions:

$$\zeta < 0: \quad v_{\zeta\zeta} + cv_\zeta - q^2 v + q^2 v_+ = 0, \quad v(0) = v_f, \ v(-\infty) = v_+, \tag{2.4a}$$

$$\zeta > 0: \quad v_{\zeta\zeta} + cv_\zeta - q^2 v + q^2 v_- = 0, \quad v(0) = v_f, \ v(\infty) = v_-, \tag{2.4b}$$

where

$$v_{\pm} = \frac{\pm 1 - a_0}{a_1 + 1/2}, \quad q^2 = a_1 + 1/2, \tag{2.5a}$$

and $v_f$ is the level of $v$ at the front position. The solutions are

$$v(\zeta) = (v_f - v_+)e^{\sigma_1 \zeta} + v_+, \quad \zeta < 0, \tag{2.6a}$$
$$v(\zeta) = (v_f - v_-)e^{\sigma_2 \zeta} + v_-, \quad \zeta > 0, \tag{2.6b}$$

with

$$\sigma_{1,2} = -c/2 \pm (c^2/4 + q^2)^{1/2}. \tag{2.7}$$

By construction, the two outer solutions for $v$ are continuous at $\zeta = 0$. Matching the derivatives of $v$ at $\zeta = 0$ gives a relation between $c$, the speed of the front, and $v_f$, the value of the $v$ field at the front position,

$$v_f = -\frac{c}{2q^2(c^2/4 + q^2)^{1/2}} - \frac{a_0}{q^2}. \tag{2.8}$$

A second relation between $v_f$ and $c$ is obtained by solving the inner problem. In the front region $u$ varies on a scale of $\mathcal{O}(\sqrt{\mu})$ but variations of $v$ are still on a scale of $\mathcal{O}(1)$. Stretching the traveling-frame coordinate according to $\chi = \zeta/\sqrt{\mu}$ we obtain from (2.1)

$$u_{\chi\chi} + \eta c u_\chi + u - u^3 - v = 0, \tag{2.9a}$$
$$v_{\chi\chi} + \sqrt{\mu} c v_\chi + \mu(u - a_1 v - a_0) = 0, \tag{2.9b}$$

where $\eta^2 = \epsilon\delta$. Setting $\mu = 0$ in (2.9b) gives the equation $v_{\chi\chi} = 0$, and we choose the solution $v = constant$. Fixing the constant, $v = v_f$, in the equation for $u$ gives a nonlinear eigenvalue problem for $c$,

$$u_{\chi\chi} + \eta c u_\chi + f(u, v_f) = 0, \tag{2.10a}$$
$$u(\mp\infty) = u_\pm(v_f), \tag{2.10b}$$

with $f(u, v_f) = u - u^3 - v_f$. The cubic function, $f$, can be rewritten as

$$f(u, v_f) = -[u - u_-(v_f)][u - u_0(v_f)][u - u_+(v_f)], \tag{2.11}$$

where $u_-(v_f) = -1 - v_f/2$, $u_0 = v_f$, and $u_+(v_f) = 1 - v_f/2$, are the linearized forms of the cubic isocline near the three solutions $u = -1, 0, 1$ respectively. The speed of the front solution of (2.10) is

$$\eta c = \frac{1}{\sqrt{2}}(u_+ - 2u_0 + u_-) = \frac{-3}{\sqrt{2}} v_f. \qquad (2.12)$$

Combining the two equations (2.8) and (2.12) we find an implicit relation for the front speed, $c$, in terms of the equation parameters $\eta, a_1$, and $a_0$,

$$\frac{\sqrt{2}}{3}\eta c = \frac{c}{2q^2\sqrt{c^2/4 + q^2}} + \frac{a_0}{q^2}. \qquad (2.13)$$

This equation was derived using the coordinate scaling (2.2). The relation for the original variables $x$ and $t$ is found by replacing $c \to c/\eta$ in (2.13):

$$c = \frac{3c}{\sqrt{2}q^2\sqrt{c^2 + 4\eta^2 q^2}} + c_\infty \qquad (2.14)$$

where $c_\infty = 3a_0/\sqrt{2}q^2$.

For the symmetric case, $a_0 = 0$ and consequently $c_\infty = 0$. Equation (2.14) then has the solution $c = 0$ representing a stationary front. This solution exists for all $\eta$ values. When $\eta < \eta_c = 3/2\sqrt{2}q^3$ two additional solutions, $c = \pm 2q\sqrt{\eta_c^2 - \eta^2}$ appear, representing counterpropagating fronts. Figure 1a displays the corresponding pitchfork bifurcation. The structures of the front solutions below and above the bifurcation are similar to those found in Ising-Bloch wall transitions [19]. For this reason we follow the terminology suggested in [8] and identify the stationary and the propagating fronts as nonequilibrium analogs of Ising and Bloch walls respectively.

For the nonsymmetric case we solved (2.14) numerically. A plot of the solutions, $c = c(\eta)$, in the $(c, \eta)$ plane yields the saddle-node bifurcation diagram shown in Figure 1b. The bifurcation point, $\eta = \eta_c$, occurs for smaller critical $\eta$ value than the symmetric case and the front that exists for $\eta > \eta_c$ is not stationary. We still refer to the two stable counterpropagating fronts beyond the bifurcation as Bloch fronts and to the single front that exists for $\eta > \eta_c$ as an Ising front.

Since $\eta^2 = \epsilon\delta$ the bifurcation point, $\eta = \eta_c$, defines a line in the $\epsilon - \delta$ plane, $\delta = \delta_F(\epsilon)$. For the symmetric case $\delta_F(\epsilon) = \frac{\eta_c}{\epsilon} = \frac{9}{8q^6\epsilon}$. For the nonsymmetric case the bifurcation line was computed numerically. Figures 2a and 2b show the bifurcation lines for the symmetric and nonsymmetric cases respectively. These results for the bifurcation line are not valid for $\delta \sim \mathcal{O}(\epsilon)$ and smaller. In that regime a different approach can be used [17].

# 3 Transverse Instability

For $\delta$ sufficiently large, planar front solutions may become unstable to transverse perturbations [20, 21, 22]. To study the transverse instabilities of the various front solutions we change from the fixed coordinate system to a coordinate system moving with the front. Let $\boldsymbol{X} = (X, Y)$ be the position vector of the front represented by the $u = 0$ contour line. The moving coordinate frame $(r, s)$ is defined by the relation

$$\boldsymbol{x} = (x, y) = \boldsymbol{X}(s, t) + r\hat{\boldsymbol{r}}(s, t), \qquad (3.1)$$

with the coordinate $s$ parameterizing the direction along the front and $\hat{\boldsymbol{r}} = \frac{Y_s \hat{\boldsymbol{x}} - X_s \hat{\boldsymbol{y}}}{\sqrt{X_s^2 + Y_s^2}}$ the unit vector normal to the front (the subscript $s$ denotes partial derivatives with respect to $s$). We assume the front radius of curvature is much larger than $l_v = \sqrt{\delta/\epsilon}$, the scale of $v$ variations across the front. We also assume the curvature varies slowly both along the front direction and in time. With these assumptions equations (1.1) assume the form

$$u_{rr} + (c_r + \kappa)u_r + u - u^3 - v = 0, \qquad (3.2a)$$
$$\delta v_{rr} + (c_r + \delta\kappa)v_r + \epsilon(u - a_1 v - a_0) = 0, \qquad (3.2b)$$

where $\kappa(s, t) = X_s Y_{ss} - Y_s X_{ss}$ is the front curvature, and $c_r(s, t) = \boldsymbol{X}_t \cdot \hat{\boldsymbol{r}}$ is the front normal velocity.

Multiplying equation (3.2b) by the factor $\Delta(s, t) = (c_r + \kappa)/(c_r + \delta\kappa)$ gives

$$u_{rr} + (c_r + \kappa)u_r + u - u^3 - v = 0, \qquad (3.3a)$$
$$\tilde{\delta} v_{rr} + (c_r + \kappa)v_r + \tilde{\epsilon}(u - a_1 v - a_0) = 0, \qquad (3.3b)$$

with $\tilde{\epsilon} = \epsilon\Delta$ and $\tilde{\delta} = \delta\Delta$. This system is exactly of the same form as equations (1.1) for a planar ($\kappa = 0$) front propagating at constant speed, $c_r + \kappa$, in the normal direction, $\hat{\boldsymbol{r}}$, except the original parameters $\epsilon$ and $\delta$ are replaced by effective parameters $\tilde{\epsilon}$ and $\tilde{\delta}$ [23]. The front bifurcation formula derived in Section 2 can now be applied to show the effects of curvature on the front velocity. Using equation (2.14) with $c$ replaced by $c_r + \kappa$ and $\eta$ by $\tilde{\eta} = \eta\Delta$ we obtain an implicit relation for the normal front velocity in terms of its curvature,

$$c_r + \kappa = \frac{3(c_r + \delta\kappa)}{\sqrt{2}q^2[(c_r + \delta\kappa)^2 + 4\eta^2 q^2]^{1/2}} + c_\infty. \qquad (3.4)$$

Equation (3.4) can be used to study the stability of the planar fronts to transverse perturbations. We look for a linear velocity curvature relation,

$$c_r = c_0 - d\kappa + \mathcal{O}(\kappa^2), \qquad (3.5)$$

valid for small curvature. Here $c_0(\eta)$ is the speed of a planar front satisfying (2.14). A positive (negative) sign of the coefficient $d$ implies stability (instability) to transverse perturbations. Inserting (3.5) into the expression for the front speed, keeping only linear terms, we find,

$$d = \frac{1}{\alpha} + (1 - \frac{1}{\alpha})\,\delta, \quad \alpha = 1 - \frac{c_0 - c_\infty}{c_0}(1 - \frac{2q^4}{9}(c_0 - c_\infty)^2). \qquad (3.6)$$

For each planar solution branch, $c_0 = c_0(\eta)$, the condition $d = 0$ defines a line in the $\epsilon - \delta$ plane where the corresponding planar front branch undergoes a transverse instability. Setting $d = 0$ for the symmetric case ($a_0 = 0$), the Ising and Bloch fronts become unstable to transverse modulations when $\delta > \delta_I(\epsilon) = \frac{8}{9}q^6\epsilon$ and $\delta > \delta_B^\pm(\epsilon) = \frac{3}{2\sqrt{2}q^3\sqrt{\epsilon}}$, respectively. The transverse instability boundary and the front bifurcation line, $\delta_F(\epsilon) = \frac{9}{8q^6\epsilon}$, are shown in Figure 2a. Figure 2b shows the transverse instability boundaries and the front bifurcation line for a typical nonsymmetric case. Note that the lines corresponding to the two Bloch fronts, denoted by $\delta_B^\pm$, are not degenerate as in the symmetric case.

# 4   Patterns in two dimensions

The two instabilities presented above provide a guide to exploring pattern types in the $\epsilon - \delta$ plane. Deep in the Ising regime there exists only one type of front and no traveling pulses or waves are expected [17]. Instead, stationary patterns may develop: ordered stripes below the transverse instability, and labyrinthine patterns above it. Far into the Bloch regime, where there is coexistence of counterpropagating fronts, traveling stripes and spiral waves appear. They are smooth below the transverse instability and develop ripples above it. The transition between these two regimes is not sharp. There exists an intermediate region, including the NIB bifurcation line, where complex spatio-temporal patterns such as replicating spots and spiral turbulence are found.

The key to understanding these complex behaviors is the highly nonlinear form of the velocity-curvature relation near the NIB bifurcation. Figure 3 shows typical solution curves of equation (3.4). The multivalued velocity-curvature relation near the NIB bifurcation (Figure 3b) unfolds to a single valued relation far in the Ising regime (Figure 3a), or folds even further to form three effectively disconnected linear branches deep in the Bloch regime (Figure 3c). *The significance of the multivalued relation near the bifurcation is that small curvature variations may drive a given branch past its endpoint and induce a transition to a different branch* [10]. Such front transitions reverse the direction of front propagation. When occurring locally they nucleate spiral-vortex pairs and may lead to spot splitting and spiral turbulence. Note that most studies of traveling waves in excitable and bistable media [11, 12, 24] have assumed a linear relation, $c_r = c_0 - d\kappa$, which is valid only deep into the Bloch regime ($\eta \ll \eta_c$) and not near the NIB bifurcation (see also [25]).

## 4.1 Labyrinthine patterns

Far into the Ising regime and beyond the transverse instability line, $\delta > \delta_I(\epsilon) = \frac{8}{9}q^6\epsilon$, front shapes meander, grow fingers, and split at the tips. This behavior can be understood using the velocity-curvature relation deep in the Ising regime, as depicted in Figure 4c. The positive slope of this relation over a wide range of curvature implies that front portions with higher curvature propagate faster, forming fingers. It also implies that the transverse instability remains effective even at the highly curved fingertips. This leads to tip splitting.

Figure 5 shows the evolution of a stripe domain in the Ising regime above the transverse instability boundary and corresponding to the velocity-curvature relation in Figure 4c. The initial stripe is perturbed transversely along the middle part. The perturbation grows, forms a meandering stripe, and then undergoes fingering and tip splitting. A final stationary labyrinth results when the pattern fills the entire domain.

Notice the final pattern in Figure 5d is connected since there were no domain fusion events during the evolution. Domain fusion is avoided by the repulsive front interactions (due to the diffusive damping of $v$ in the region between approaching fronts [17]). Closer to the front bifurcation the front speeds are higher (see Figure 1b) and the repulsive interactions may not be strong enough to prevent fusion. As a result the eventual stationary pattern may contain disconnected domains.

Similar labyrinthine patterns have been observed in the bistable FIS reaction [5]. Our interpretation is that these patterns occur in the Ising regime where the single front structure corresponds to a high pH state invading a low pH state.

## 4.2 Single spot dynamics

Closer to the NIB bifurcation the nonlinearity of the velocity-curvature relation becomes important. Consider an up state disk expanding radially outward. Depending on the system parameters several scenarios for evolution are possible. Deep enough into the Ising regime, where the velocity-curvature relation is still single valued (see Figure 4a), a stationary disk solution exists. The disk has radius $1/\kappa_0$, where $c_r(\kappa_0) = 0$, and is stable to uniform expansions and contractions because the velocity-curvature relation has positive slope at $c_r = 0$ (it might be unstable, however, to transverse perturbations [21]).

Still closer to the front bifurcation (but in the Ising regime) the velocity-curvature relation becomes multivalued and the slope at $c_r = 0$ negative as illustrated in Figure 4b. The stationary disk is no longer stable to expansions and contractions and a breathing disk solution appears [21]. To understand this breathing motion, note first that the boundary of an expanding disk corresponds to a front lying on the upper branch in Figure 4b. As the disk expands the front curvature decreases. When the curvature falls below the value where the upper branch terminates a transition to the lower branch takes place. The disk stops expanding and starts contracting. The curvature increases until the endpoint of the lower branch is reached and a transition back to the upper branch occurs. As a result the disk stops contracting and begins expanding again. These oscillations are similar to those found in one-dimensional domains [17, 18, 26] with front interactions playing the role of curvature in inducing front transitions [10].

To verify these expectations we studied numerically equations (1.1) in polar coordinates assuming circularly symmetric front solutions (to avoid transverse instabilities). Such solutions satisfy

$$u_t = u_{rr} + \frac{1}{r}u_r + u - u^3 - v, \qquad (4.1a)$$

$$v_t = \delta v_{rr} + \frac{\delta}{r}v_r + \epsilon(u - a_1 v - a_0). \qquad (4.1b)$$

They represent circular fronts with curvatures $\kappa(t) = 1/r_0(t)$, where $r_0(t)$ solves $u(r_0, t) = 0$. Figures 6a and 6b show the curvatures of up state disks as functions of time for parameters pertaining to Figures 4a and 4b respectively. A single valued (multivalued) velocity-curvature relation leads to a stationary (oscillatory) asymptotic state.

In fully two-dimensional systems oscillating disks might be unstable to non-circularly symmetric perturbations. Consider an expanding disk perturbed to an oval shaped domain as shown in Figure 7a. The parameters chosen pertain to a velocity-curvature similar to the one in Figure 4b. As the domain expands the flatter portions of its boundary are the first to reach the end of the upper branch and undergo front transitions. These portions then propagate towards one another, annihilate, and split the domain as shown in Figures 7b and 7c. The crossing points of the $u = 0$ and $v = 0$ contour lines indicate the cores of spiral vortices. Note that the splitting process involves the creation and the subsequent annihilation of two spiral-vortex pairs. A successive splitting is shown in Figure 7d. The asymptotic state in this case is a disordered stationary pattern with many disconnected domains. Remnants of the unstable breathing motion are often seen when the split spots first contract, then approach a minimum size and start expanding. Both spot splitting and the persistence of small spots have been observed recently be Lee *et al.* in the FIS reaction [6] [27].

### 4.3 Spiral turbulence

Further approach to the NIB bifurcation results in disordered dynamic patterns where spiral vortices nucleate and annihilate repeatedly. We refer to such a state as spiral turbulence. The nucleation of spiral-vortex pairs results from local front transitions, very much like in spot splitting except the mechanisms that drive the transitions are different and depend on the system parameters. For $\delta$ sufficiently large, the transverse instability plays a dominant role in inducing such transitions [28]. Figures 8 and 9 show spiral turbulence for smaller $\delta$ values. In this parameter regime front interactions are the major driving force. Figures $4f$ and $4g$ show the corresponding velocity-curvature relations. In both figures the upper branches terminate at positive curvature values. Processes reducing the curvature of fronts on these branches past the endpoints cause transitions to the lower front branches. As we have seen in the previous section, single noncircularly symmetric domains may undergo such transitions and split. In the presence of nearby domains, however, repulsive front interactions accelerate domain splitting by flattening out approaching curved fronts. Frames $d, e, f$ of Figures 8 and 9 show local front transitions and splitting driven by front interactions (see the regions indicated by the arrows). These processes are strikingly similar to those observed in the FIS reaction [27]. Front interactions may also cause spiral-vortex nucleation and splitting by directly inducing front transitions. Reflection of one-dimensional fronts provides an example of front interactions leading to front transitions [17]. The spiral-vortex nucleations in Figures 8 and 9 are likely to result from both mechanisms.

The patterns in Figure 8 differ from those in Figure 9 in a few respects. The initial conditions in both simulations are the same but the subsequent spiral breakup is different. In Figure 8 most of the spiral wave disappears as the weakly curved front away from the core undergoes a front transition. In Figure 9 the weakly curved part of the spiral survives longer until front interactions or interactions with the boundaries come into play. This is because the upper branch of the the velocity-curvature relation pertaining to Figure 9 terminates at a lower curvature value than that of Figure 8 (see Figures $4f$ and $4g$). Another difference is the prevalence of more spots in the patterns of Figure 8. This is partly because domain fusions are avoided in Figure 8 but occasionally take place in Figure 9 (closer to the NIB bifurcation, front speeds are higher and domain fusions are more likely).

# 5 Numerical Methods

Numerical simulations were performed by integrating the system (1.1) with Neumann boundary conditions. The spatial derivatives were computed by 4th order finite difference approximations on a uniform $400 \times 400$ grid over a domain of $0 \leq x \leq 400$, $0 \leq y \leq 400$. For typical equation parameters the $400 \times 400$ resolution resulted in about 6 points across the narrow front region in $u$. Doubling the resolution slightly changed the location of the front bifurcation but did not affect the qualitative results. For the turbulent patterns (Figures 8 and 9) the time integration was performed with a second order Adams method. In the simulations for the labyrinthine patterns and spot dynamics (Figures 5 and 7) the explicit Adams method was replaced with an implicit second order formula for efficiency of integration on long time scales [29]. For the circularly symmetric simulations a one-dimensional version of the 4th order finite difference discretization was used with a variable order, variable step size, implicit time integration scheme.

# 6 Conclusion

The Labyrinthine patterns found in the Ising regime have also been studied by Petrich and Goldstein [30] using a nonlocal interface model. The model was derived for parameters far into the Ising regime where the reaction-diffusion equations reduce to a gradient system. We expect the qualitative predictions of this model to hold closer to the NIB bifurcation under two conditions: domain fusion, which changes the topology of the front curve, does not take place, and nongradient effects, such as front transitions, do not appear.

Spot splitting has been studied numerically by Pearson [31] and analytically, in one space dimension, by Reynolds *et al.* [32]. No one-dimensional analog of spot splitting exists in the present model, at least for the parameter regime where splitting in two dimensions has been observed. The splitting studied here is a purely two-dimensional effect where curvature and front interactions play key roles. We believe these factors also have important roles in the two-dimensional spot splitting simulations by Pearson.

The breakup of spiral waves studied here results from local front transitions that become feasible near the NIB bifurcation. The transitions can be induced by quite general perturbations, either extrinsic or intrinsic. In [10] we studied front transitions caused by external advective fields. The transverse instability also can lead to local front transitions by creating negatively curved front portions [28]. Here the important factors are front interactions that act either directly, or indirectly (by flattening curved fronts). It would be interesting to investigate the relevance of the above mechanism to recent studies of spiral turbulence in surface reactions [33] and in cardiac tissue models [34, 35, 36, 37, 38].

Many of the observations made here have also been found in the FIS reaction. These include labyrinthine patterns [5], spot splitting [6], and front transitions induced by front interactions [27]. Further comparative investigation should include experimental testing for the existence of a NIB bifurcation, and examination of the experimental observations in relation to the location of the bifurcation.

We expect similar complex spatio-temporal patterns to be found in other systems exhibiting NIB bifurcations. Periodically forced oscillatory systems [8] and liquid crystals subjected to rotating magnetic fields might be good candidates [39, 40].

# Acknowledgments

This work was supported in part by the Arizona Center for Mathematical Sciences (ACMS), sponsored by AFOSR. One of us (A.H.) acknowledges the support of the Computational Science Graduate Fellowship Program of the Office of Scientific Computing in the Department of Energy.

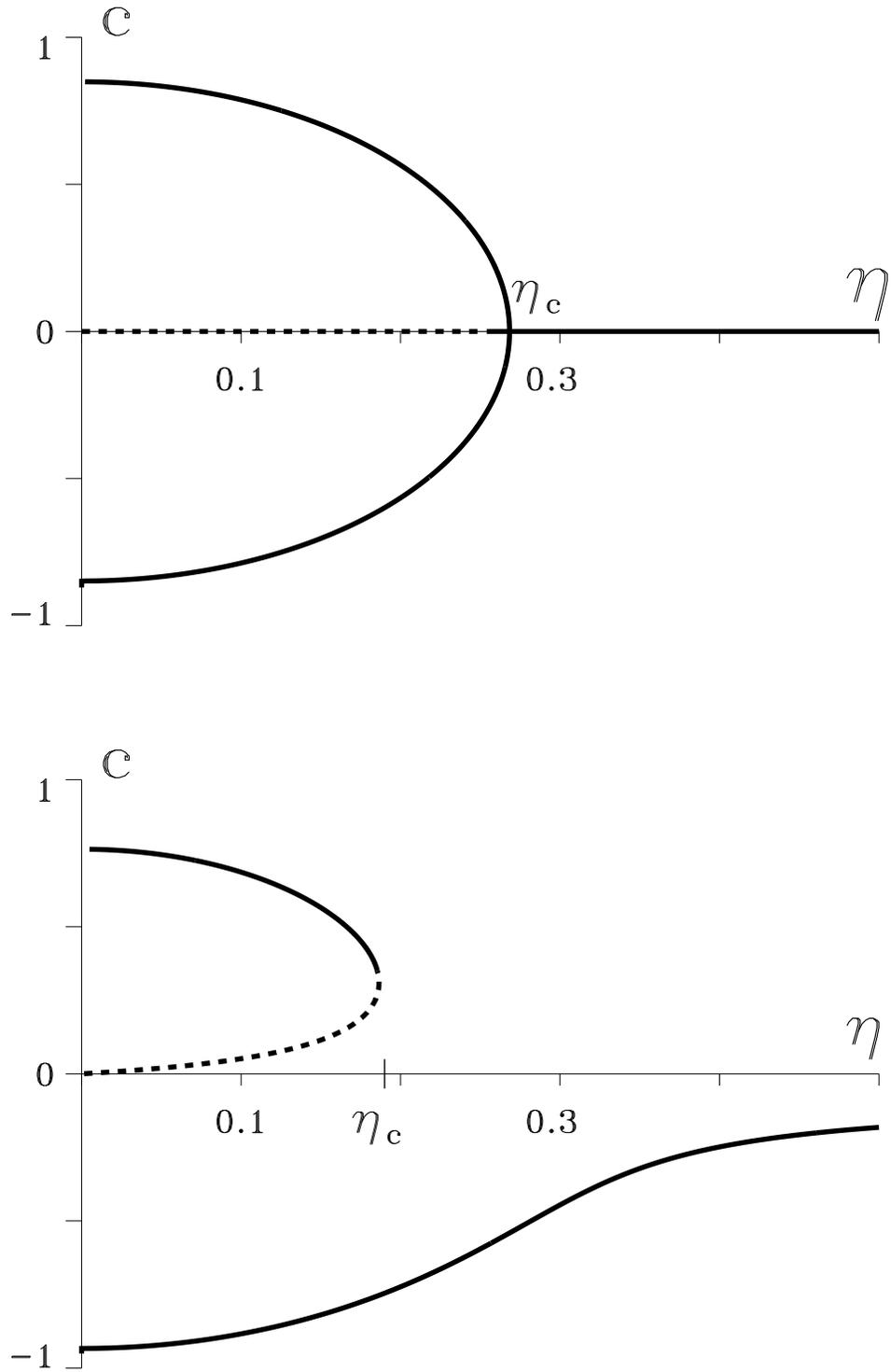

Figure 1: The NIB front bifurcation in the $(c, \eta)$ plane. The solid (broken) line represents a stable (unstable) branch of front solutions. (top) The symmetric case, $a_0 = 0$. (bottom) The nonsymmetric case, $a_0 = -0.1$.

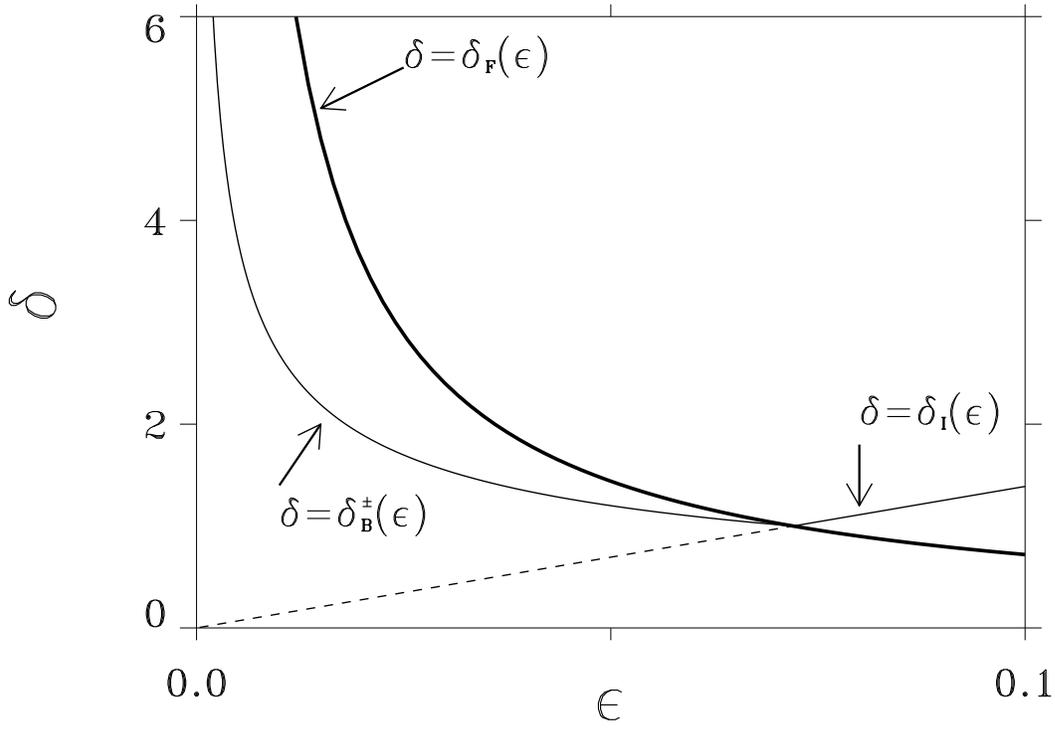

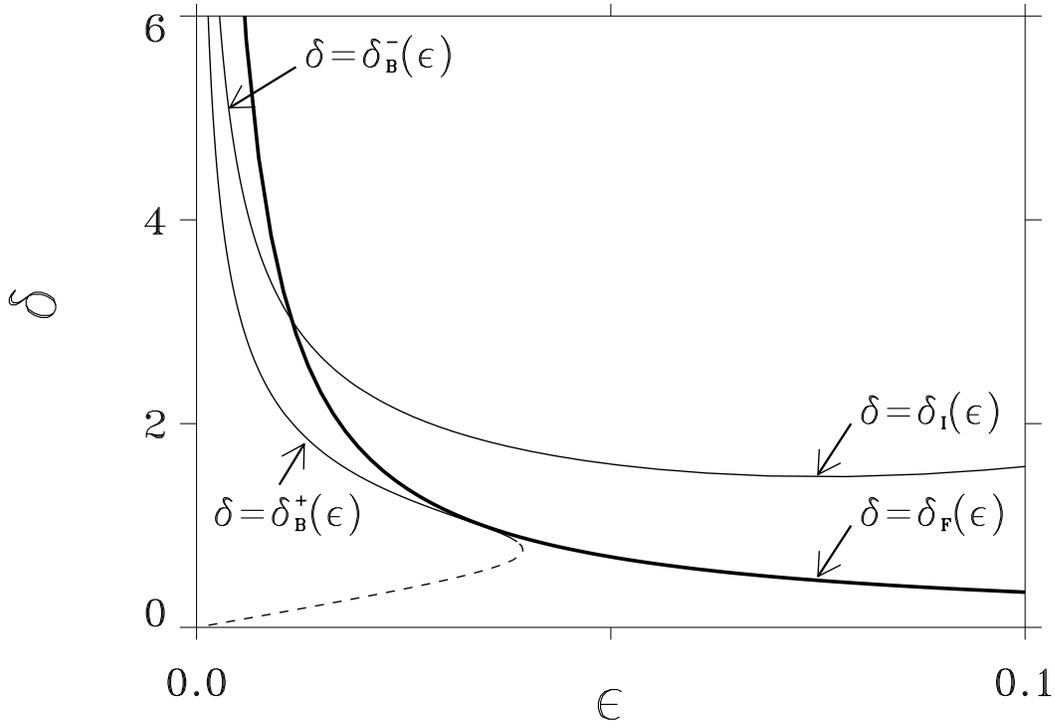

Figure 2: The NIB bifurcation and the transverse instability lines in the $\epsilon-\delta$ parameter plane for (top) the symmetric case ($a_0 = 0$), and (bottom) the nonsymmetric case ($a_0 = -0.1$). The front bifurcation, $\delta = \delta_F(\epsilon)$, is indicated by the thick line. The transverse instabilities are indicated by the thin lines, $\delta = \delta_I(\epsilon)$ for Ising fronts, and $\delta = \delta_B^{\pm}(\epsilon)$ for Bloch fronts.

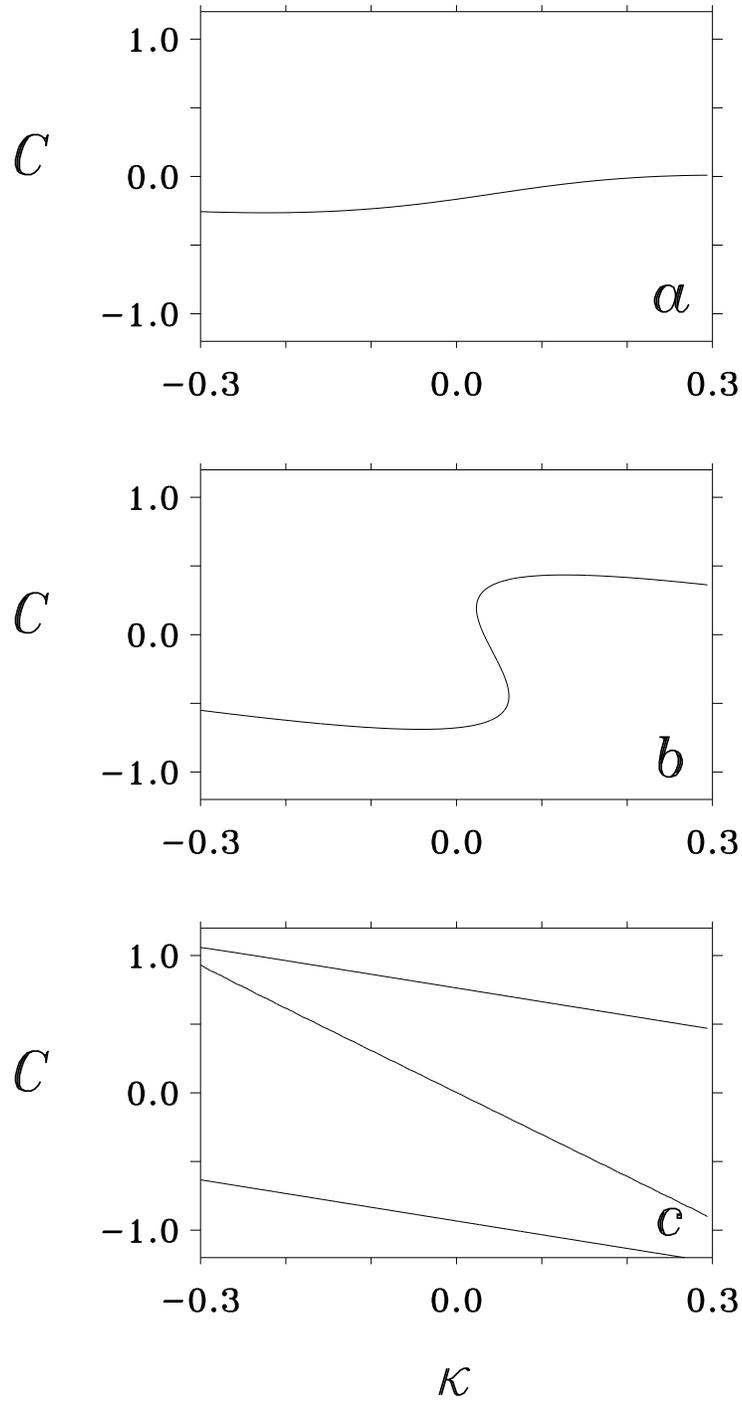

Figure 3: Front velocity, $c$, vs. curvature, $\kappa$. (a) Deep in the Ising regime. (b) Near the NIB bifurcation. (c) Deep in the Bloch regime.

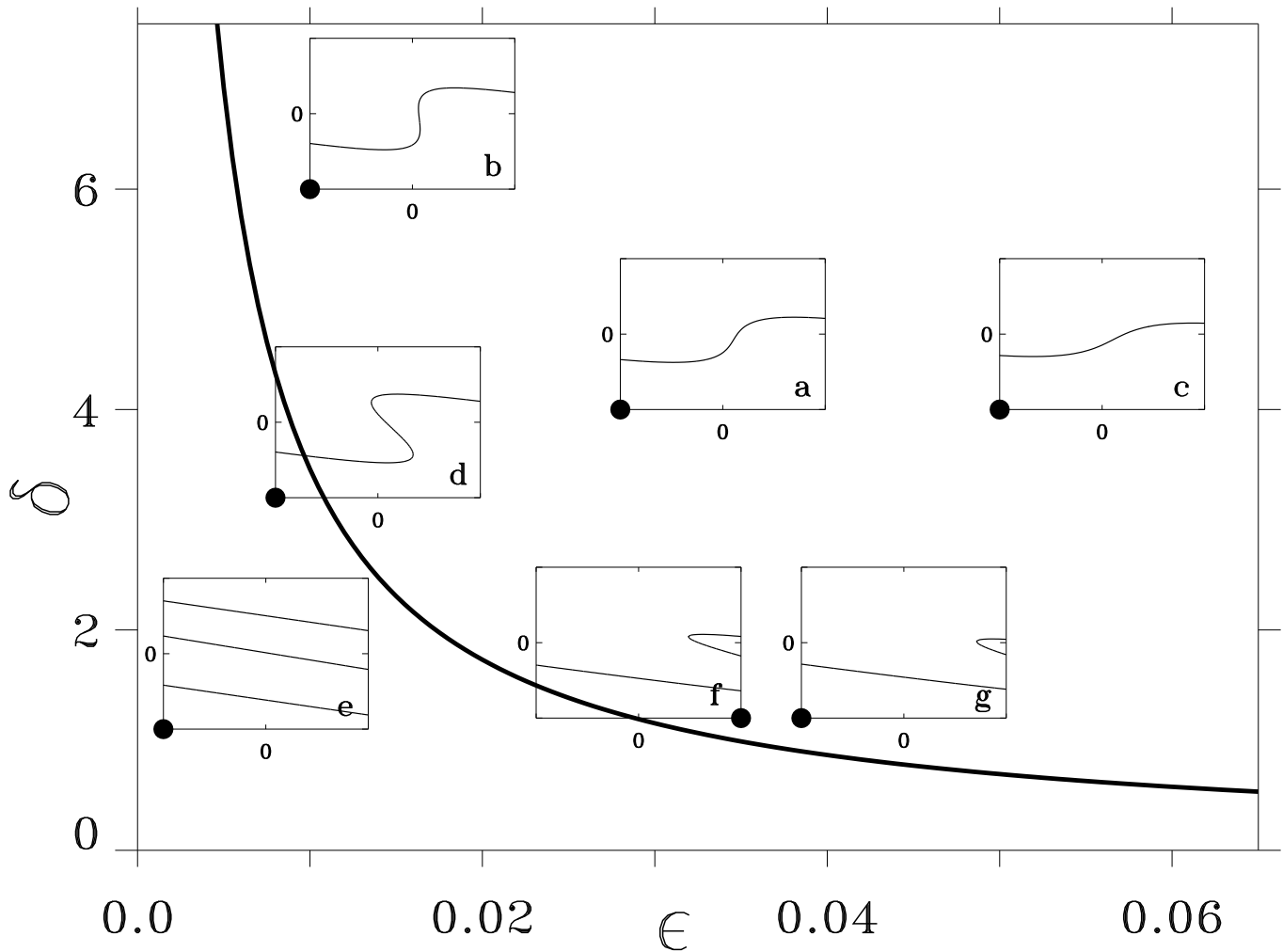

Figure 4: A map of the velocity vs. curvature relation in the $\epsilon - \delta$ plane. The thick line represents the front bifurcation for $a_0 = -0.1$ and the insets display $c$ vs. $\kappa$ for the point indicated by the solid circle. The axis scales for the insets are the same as in Figure 3.

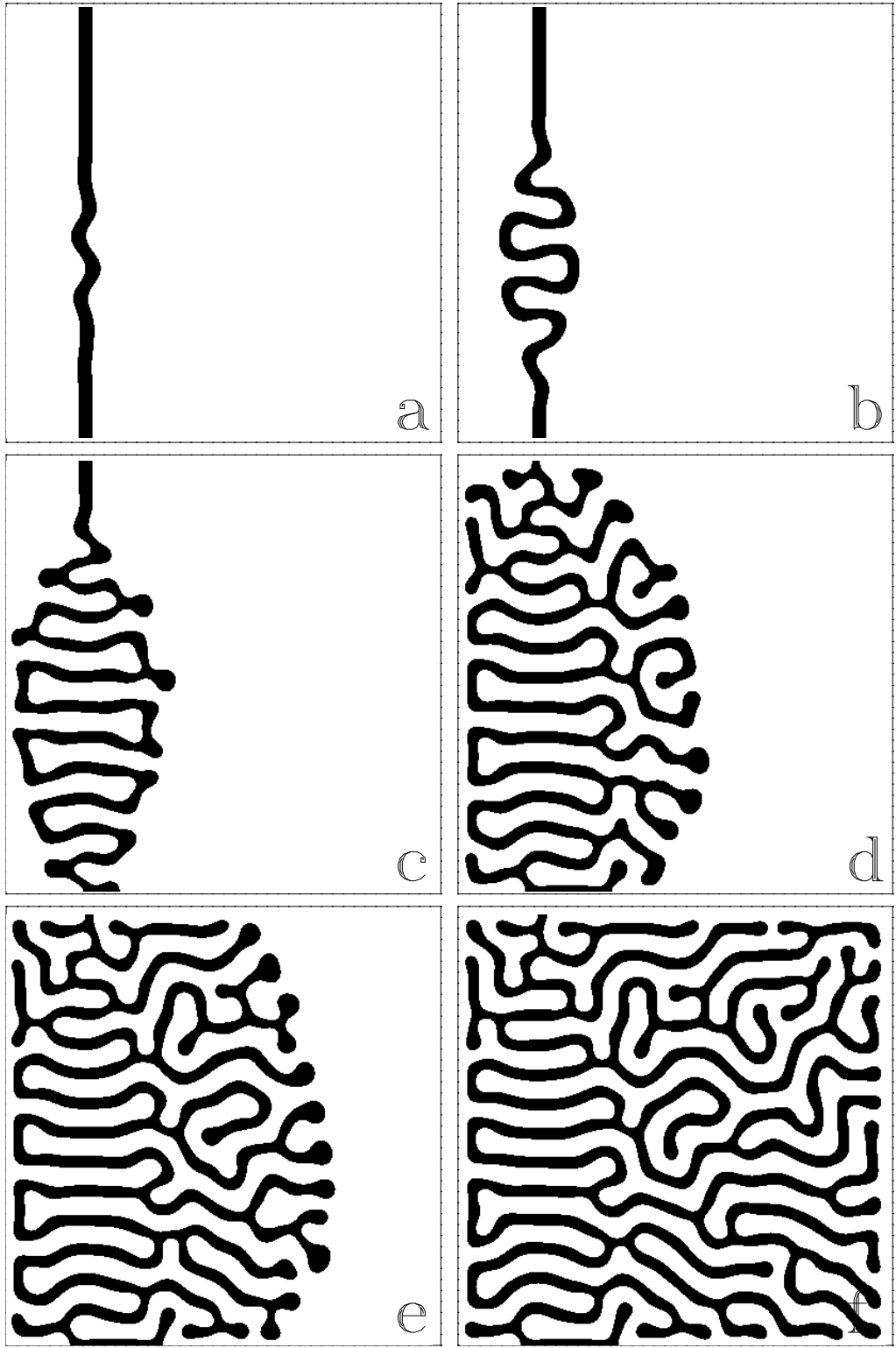

Figure 5: The evolution of an initially perturbed stationary stripe in the Ising regime above the transverse instability line (see Figure 4c). The dark and light regions correspond to the up and down states respectively. The frames $a, b, c, d, e, f$ pertain to times $t = 100, 525, 1100, 1900, 2675, 5000$. The computational parameters are $a_1 = 2.0$, $a_0 = -0.1$, $\epsilon = 0.05$, $\delta = 4.0$ on a domain of $0 \leq x \leq 400$, $0 \leq y \leq 400$.

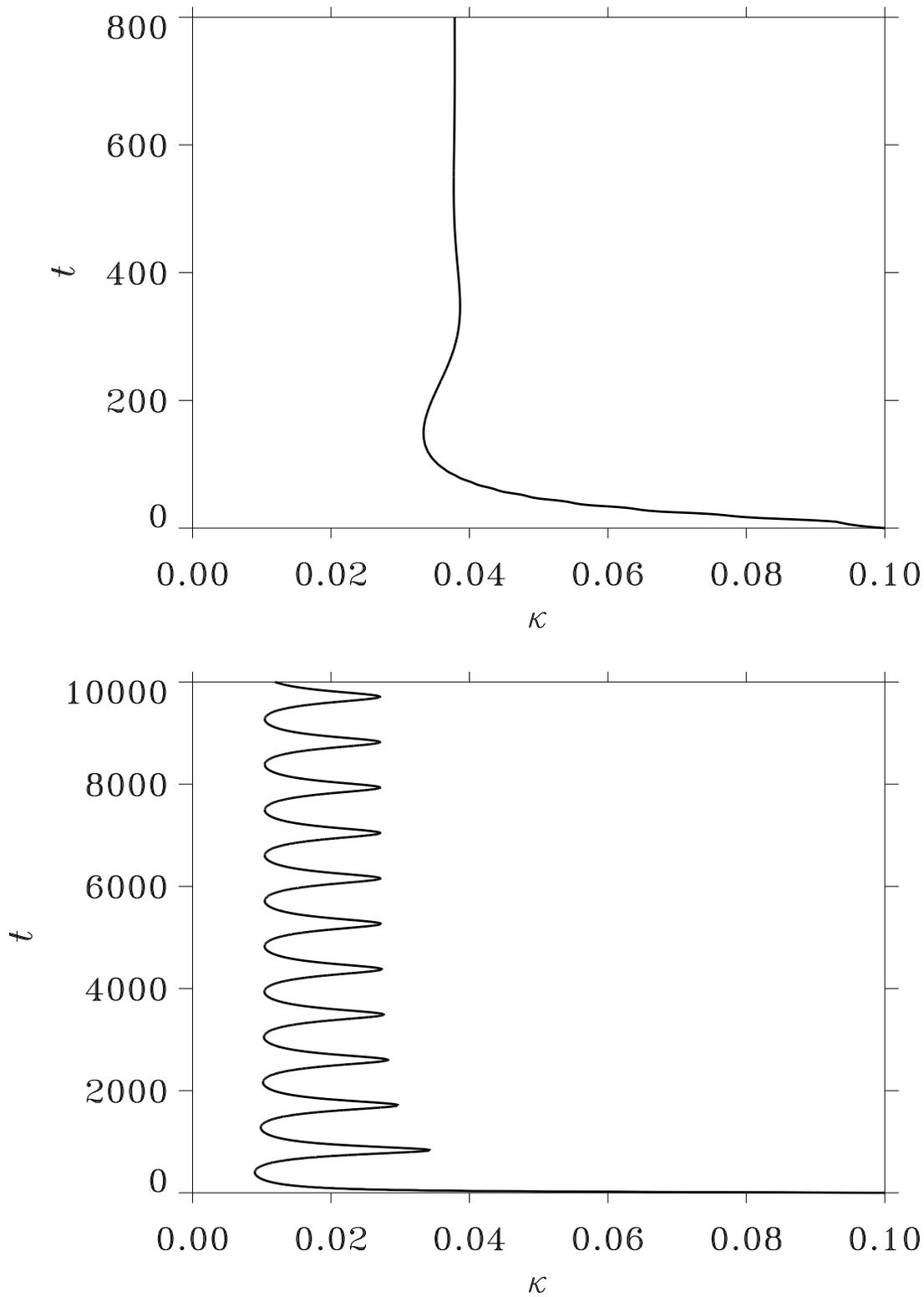

Figure 6: Curvature, $\kappa$, vs. time for a disk shaped domain. (top) Far in the Ising regime (Figure 4a) a stationary disk pattern is reached. Computational parameters: $a_1 = 2.0$, $a_0 = -0.1$, $\epsilon = 0.028$, $\delta = 4.0$. (bottom) Near the NIB bifurcation (Figure 4b) oscillations set in. Computational parameters: $a_1 = 2.0$, $a_0 = -0.1$, $\epsilon = 0.01$, $\delta = 6.0$.

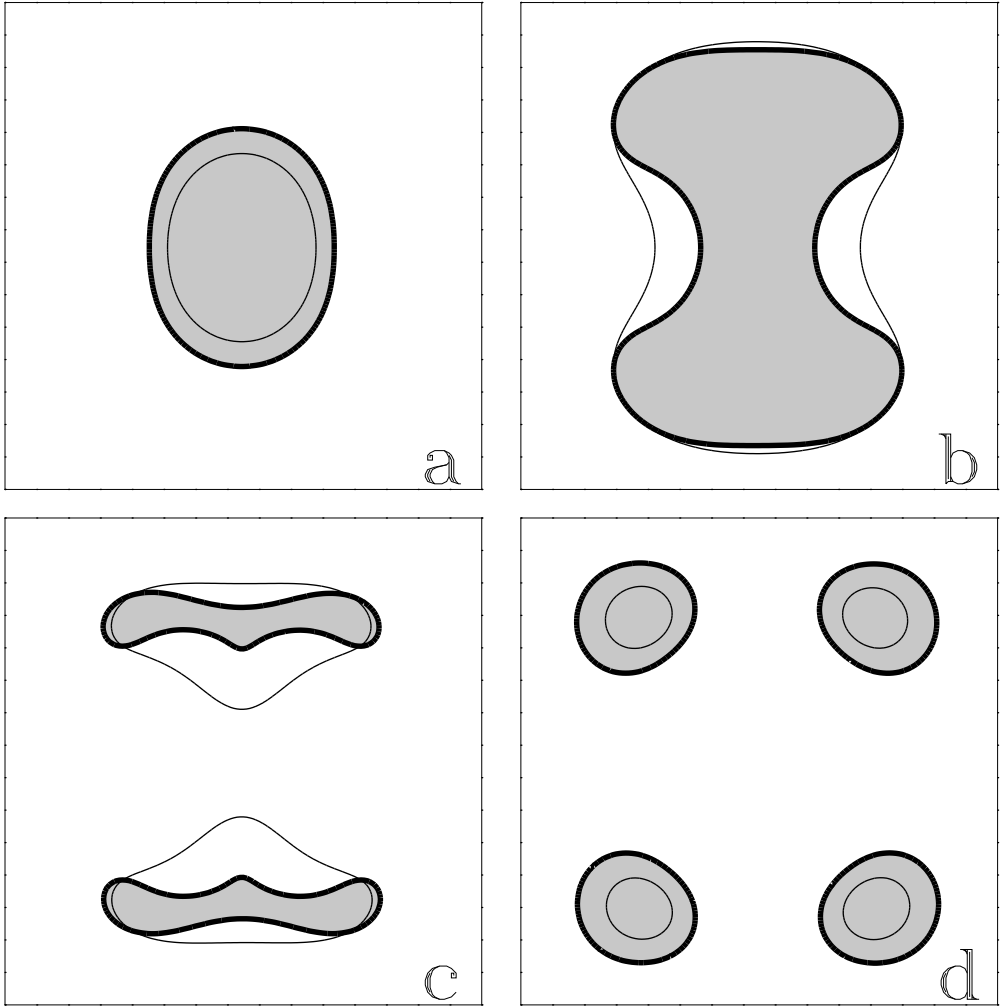

Figure 7: Domain splitting of an oval shaped domain. The shaded (light) region corresponds to the up (down) state and the thick (thin) line represents the contour of the $u = 0$ ($v = 0$) field. Frames $a, b, c, d$ pertain to times $t = 80, 240, 280, 340$. Local front transitions occur at the flatter portions of the front. They are accompanied by nucleation of vortex pairs, and followed by domain splitting. Computational parameters: $a_1 = 2.0$, $a_0 = -0.15$, $\epsilon = 0.014, \delta = 3.5$.

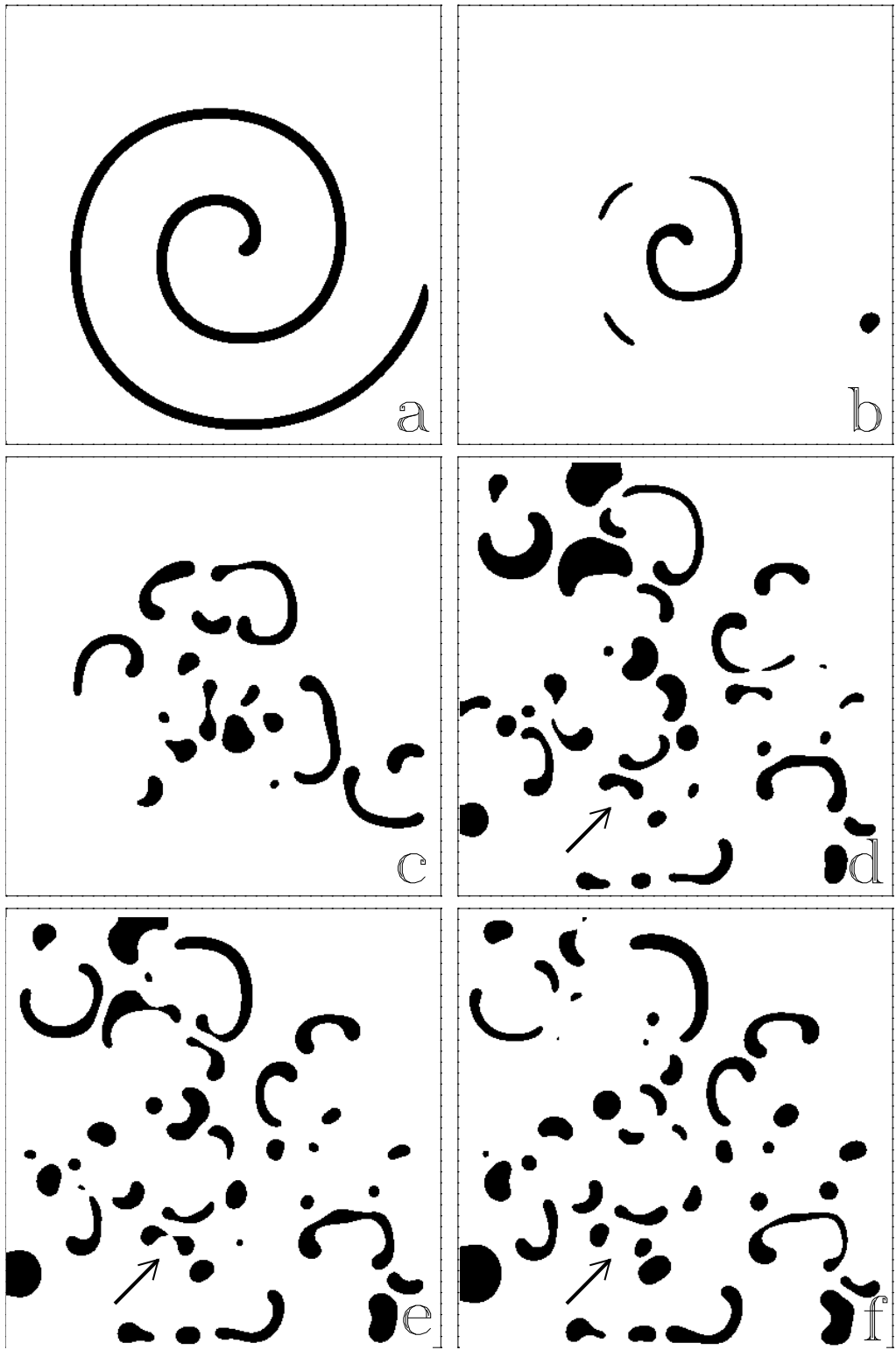

Figure 8: Spiral breakup for the velocity-curvature relation of Figure 4$g$. The frames $a, b, c, d, e, f$ represent the solution at times $t = 60, 640, 1540, 3760, 3780, 3800$ respectively. The computational parameters are $a_1 = 2.0$, $a_0 = -0.10$, $\epsilon = 0.0375$, $\delta = 1.2$ on a domain of $0 \leq x \leq 400$, $0 \leq y \leq 400$. For a more detailed description see the text.

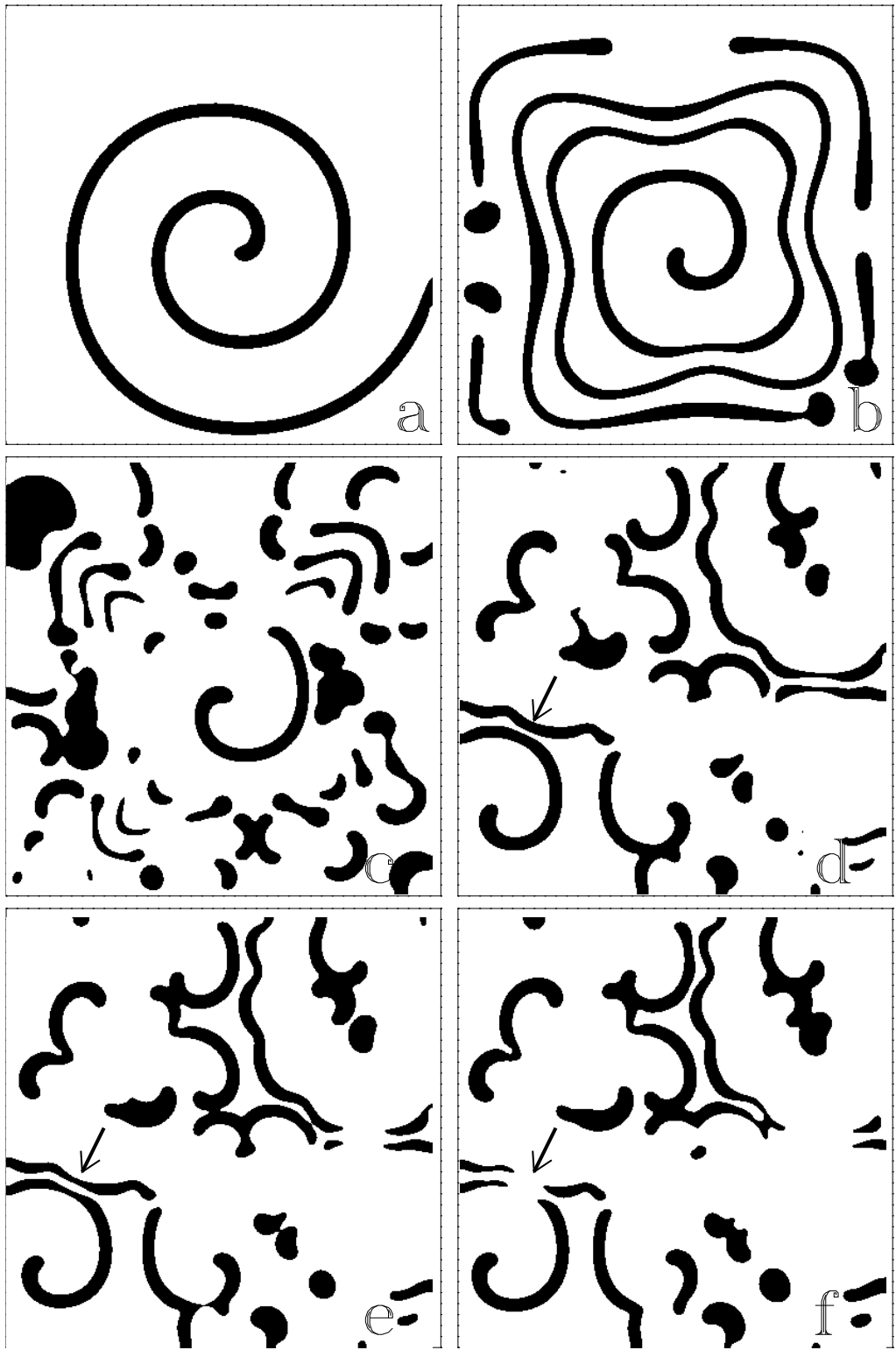

Figure 9: Spiral breakup for the velocity-curvature relation of Figure 4$f$. The frames $a, b, c, d, e, f$ represent the solution at times $t = 80, 620, 900, 1890, 1900, 1910$ respectively. The computational parameters are the same as Figure 8 with $\epsilon = 0.035$. For a more detailed description see the text.